# Ion acceleration with few cycle relativistic laser pulses from foil targets

Sargis Ter-Avetisyan[1], Parvin Varmazyar[1], Prashant K. Singh[1], Joon-Gon Son[1], Miklos Fule[1], Valery Yu. Bychenkov[2,3], Balazs Farkas[4], Kwinten Nelissen[4], Sudipta Mondal[4], Daniel Papp[4], Adam Börzsönyi[4], Janos Csontos[4], Zsolt Lécz[1,4], Tamas Somoskői[4], Laszló Tóth[4], Szabolcs Tóth[4], Velyhan Andriy[5], Daniele Margarone[5], Ales Necas[6], Toshiki Tajima[6], Gerard Mourou[6], Gabor Szabó[4,8], and Karoly Osvay[1,8]

[1]*National Laser-Initiated Transmutation Laboratory, University of Szeged, 6720 Szeged, Hungary*
[2]*P. N. Lebedev Physics Institute, Russian Academy of Sciences, 119991 Moscow, Russia*
[3]*Center for Fundamental and Applied Research, VNIIA, ROSATOM, Moscow 127055, Russia*
[4] *ELI-ALPS, ELI-HU Non-Profit Ltd., Szeged, Hungary*
[5] *ELI–Beamlines Center, Institute of Physics, Czech Academy of Sciences, 252 41 Dolní Břežany, Czech Republic*
[6] *TAE Technologies Ltd, Pauling, Foothill Ranch, CA, USA*
[7] *IZEST, Ecole Polytechnique, Palaiseau, France*
[8]*Dept. Optics and Quantum Electronics, University of Szeged, 6720 Szeged, Hungary*

Ion acceleration resulting from the interaction of 11 fs laser pulses of ∼35 mJ energy with ultrahigh contrast ($<10^{-10}$), and $10^{19}$ W/cm$^2$ peak intensity with foil targets made of various materials and thicknesses at normal (0°) and 45° laser incidence is investigated. The maximum energy of the protons accelerated from both the rear and front sides of the target was above 1 MeV. A conversion efficiency from laser pulse energy to proton beam is estimated to be as high as ∼1.4 % at 45° laser incidence using a 51 nm-thick *Al* target. The excellent laser contrast indicates the predominance of vacuum heating via the Brunel's effect as an absorption mechanism involving a tiny pre-plasma of natural origin due to the Gaussian temporal laser pulse shape. Experimental results are in reasonable agreement with theoretical estimates where proton acceleration from the target rear into the forward direction is well explained by a TNSA-like mechanism, while proton acceleration from the target front into the backward direction can be explained by the formation of a charged cavity in a tiny pre-plasma. The exploding Coulomb field from the charged cavity also serves as a source for forward-accelerated ions at thick targets.

## I. INTRODUCTION

At the dawn of the field, it was possible to accelerate protons and ions above MeV energies with large scale laser systems firing only a few times in an hour. So far, the highest laser-to-ion conversion efficiency has been demonstrated with sub-ps laser pulses [1]. Laser systems with pulse durations of a few tens of femtoseconds provide the necessary peak intensity but with lower pulse energy and higher repetition rate [2]. Various experiments [3-7] and theoretical studies [8-13] investigated the acceleration mechanisms on thin targets and optimised the ion yield. Such laser-driven ion beams have demonstrated their credibility for applications in a wide range of areas, e.g., probing of fields in plasmas [14], generation of directional neutron sources [15], isochoric heating of solid matter [16], and radiobiology [17-18]. Recent developments in sub-50 fs laser systems made possible to optimise ion acceleration [19-20], and start building beamlines dedicated especially for medical use of laser-accelerated ions [21-23].

Currently, the advances in laser technology have led to the generation of laser pulses with durations of only a few optical cycles. Due to the very short pulse duration as well as the extreme beam quality, it has become possible to reach the relativistic intensity in the focus with a few tens of mJ pulse energy. At such low energy, few-cycle laser systems are capable of operating at kHz repetition rate over 24/7 with high reliability [24], offering an opportunity to generate stable ion beams [25] and, possibly, gamma-rays. These could be highly beneficial for medicine, industry, science, and home-land security [26]. For example, high repetition rate ion sources can serve as input beams of diverse accelerator systems, meeting the requirements of high current, low and high charge state, and low emittance [27]. Deuterons accelerated to energies above 0.5 MeV at kHz repetition rate can generate a large flux of neutrons per second via D(d, n) reaction. Radioisotope production with kHz repetition rate is another potential application for nuclear pharmacology [28,29]. Apart from an early work [30] and a recent study [31] with somewhat lower intensity on a bulk target, experimental results on ion acceleration with few-cycle relativistic pulses are yet to be disseminated.

In this paper we discuss our recent experimental findings with the use of ultra-high contrast, few cycle laser pulses of relativistic intensity on diverse target materials, thicknesses, and angle of incidence. The measurements show new features in the ion acceleration scenario. The maximum energy for protons was measured just above 1 MeV, being far less affected by target material (metal or dielectric) and thickness, than expected from present ion acceleration theories [8-13]. At a laser incidence angle of 45°, the properties of proton emissions from the rear and the front of the target surface were similar. The cut-off energy was, however, about half of that of at normal laser incidence on the target. We also provide quantitative explanations for the novel features observed in the given ion acceleration scenario.

## II. EXPERIMENTAL SETUP

In the experiments the charged particles were generated by irradiating foil targets with the SYLOS Experimental Alignment (SEA) Laser at the ELI-ALPS facility in Szeged, Hungary [24]. In the experiments (Fig 1.a), $p$-polarised, 11 fs laser pulses of ~35 mJ energy were focused using a f/2.5 off axis parabolic mirror to a focal spot of ~3 μm full width half maximum (FWHM). The intensity of $I_0 \sim 10^{19}$ W/cm$^2$ in the focus was inferred from pulse energy, pulse duration, and focal spot size measurements (containing 35% of energy). It is worth mentioning that the measured ultrahigh contrast <10$^{-10}$ (Fig. 1.b) is still conservative due to the bandwidth limitation of the third order cross-correlator (Sequioa by Amplitude Tech). Since the standard thicknesses of the non-linear crystals support a bandwidth half of SEA laser pulse, the laser pulse peak intensity and hence, the temporal contrast is underestimated. As a consequence, the contrast is expected to be ~8 times higher than that is shown in Fig.1b).

The foil targets consisted of different materials: carbon ($C$), diamond-like-carbon (DLC), formvar ($[CH_2CH(OH)]_n$), and aluminium ($Al$). The targets were irradiated at an incidence angle of 0° and 45° with respect to the target normal. The target thickness was varying in the range 5 nm - 9 μm. A target alignment system consisting of an infinity-corrected long working distance objective located along the axis of the incident laser beam is used to visualize the laser focal spot and position the target in the same focal spot. The targets were positioned with an accuracy of five micrometres at the laser focal plane [32] which ensured high reproducibility and reliability of the experimental data.

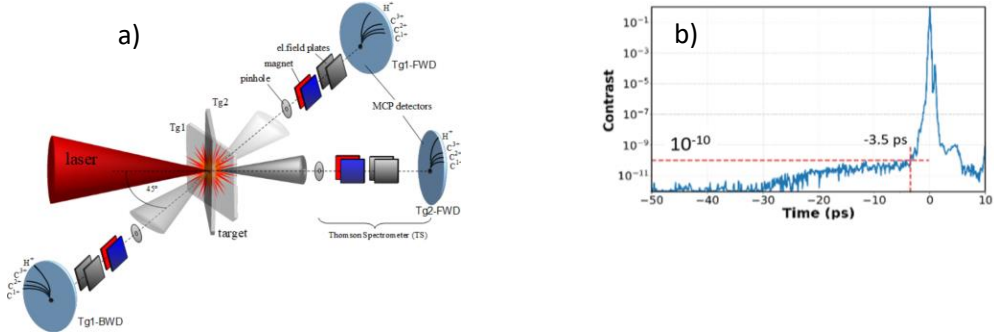

Fig. 1. a) The experimental arrangement. Laser incidence angle on target Tg1 and Tg2 is 45° and 0°, with respect to target normal direction, respectively. Thomson spectrometers are located in a rear or front surface target normal direction: from Tg1 the ions are measured in Tg1-FWD and in Tg1-BWD directions and from Tg2 in Tg2-FWD direction. b) The measured temporal contrast of the laser pulse.

The ion spectra were measured in a single laser shot with a calibrated Thomson spectrometer [33] located in a rear or front surface target normal direction, later called forward (FWD) and backward (BWD) directions. The collection of ions under the solid angle of 30 nsr ensured the measurement of well resolved ion spectra within the energy range of the spectrometer $E_i/Z \approx 0.025 \div 2$ MeV/nucleon. Typically, a magnetic field of (0.27±0.01) T and an electric field up to 3 kV/cm were applied. The energy resolution of the spectrometer at high energies (above 0.5 MeV) was $\Delta E/E_i \sim 5\%$. The ions were detected by an MCP detector coupled to a phosphor screen. Our particular interest is the ion spectrum, from which the maximum ion energy as well as the number of ions can be obtained. Typical measured ion spectra accelerated are shown in Fig 2, along with the corresponding evaluated proton spectrum.

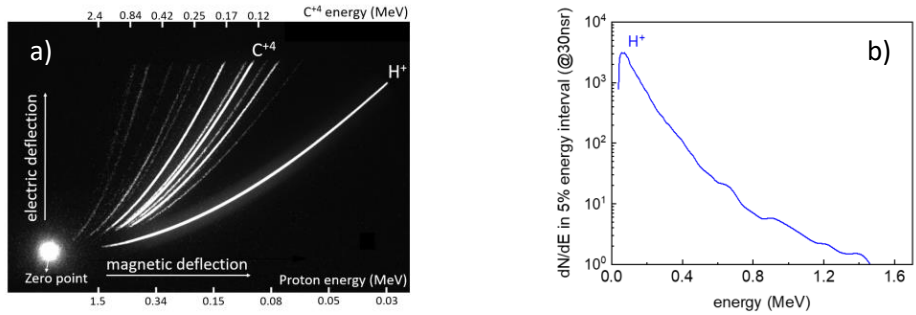

Fig. 2. (a) A typical CCD picture of measured ion spectra from a 26 nm formvar target, as an example, and (b) the evaluated proton spectrum (the background is substracted).

III. EXPERIMENTAL RESULTS

At an incidence angle of 45° on the target, the maximum proton energies observed in FWD and BWD versus target thickness and materials are shown in Fig. 3 a) and b), respectively. The noticeable feature here is that the acquired maximum energy of protons from targets with thicknesses from 2 µm down to 20 nm only slightly depends on target thickness. In FWD (Fig. 3a)) the measured maximum proton energy (1.2±0.3 MeV) is found around 200 nm, slightly decreasing to about 0.8 MeV for the thinnest (5 nm) and the thickest (9 µm) targets. In BWD (Fig. 3b)) the proton energies are somewhat lower (compared to that of in FWD). In both cases the formvar targets show a higher proton cut-off energy compared to other target materials. In FWD, the proton energy from formvar target is about 1.2 times higher than in BWD, while for the other target materials the difference is almost twice. It worth mentioning that symmetric proton emission in FWD and BWD directions was observed in Ref. 34 at ~6 times longer laser pulse and ~18 times more pulse energy as compare to our case, which is very far from the few cycle regime. Hence, the asymmetric behavior we observed suggests some peculiar feature of proton acceleration in the few-cycle regime. The implications on the acceleration conditions, including preplasma effects on the front and back of the target are discussed in a further chapter of this paper.

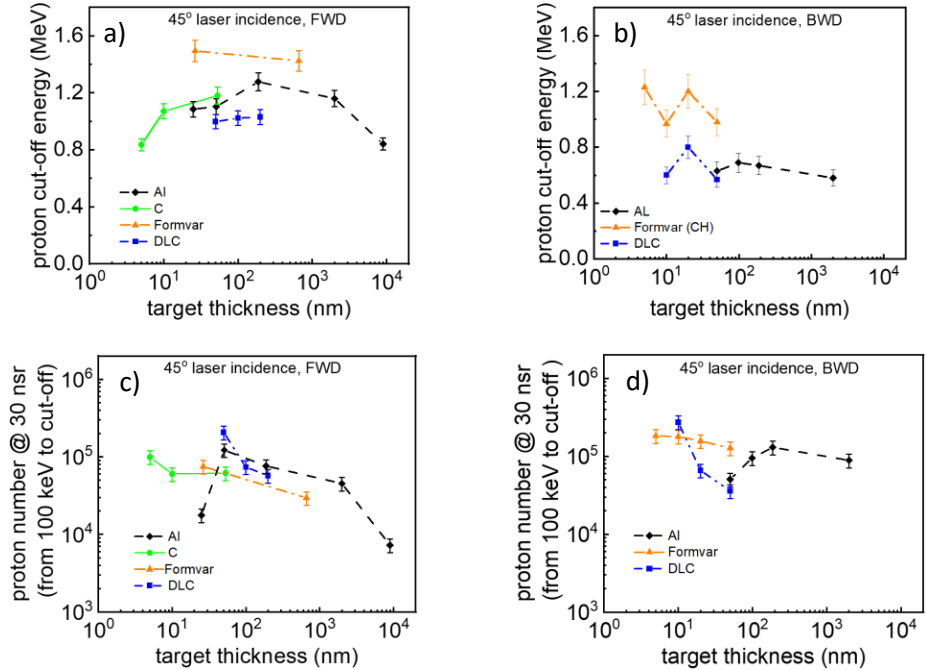

Fig. 3. (a), (b) the maximum energy of protons and (c), (d) highest number of protons versus thickness of the target emitted in forward (a), (c) ) and backward (b), (d)) directions are shown at 45∘ laser incidence angle on different target materials.

In Fig. 3 c) and d), the number of protons emitted in a 30 nsr solid angle along the FWD and BWD directions are shown. The number of particles is integrated above 100 keV up to cut-off energy. It is worth noting that it was not always the spectrum with the highest cut-off energy that showed the highest particle yield. In Fig. 3 c) and d) the measured highest proton yield is about $10^5$ protons within a factor of $2 - 4$, while the target thickness varies several orders of magnitude.

At a 0° angle laser incidence on the target the maximum proton energies versus target thickness observed in FWD direction are shown in Fig. 4a) for different foil materials. Here, compared to a 45° laser incidence (Fig. 3) the energies are over a factor of two lower, while the integrated numbers of protons is an order of magnitude lower, except for the formvar target. However, it shows a similar dependence of cut-off energy to 45° laser incidence with a slight decrease at the thinnest and the thickest targets and the similar tendency to be almost independent of the target material. Such a weak dependence of proton cut-off energy and target thickness may indicate that in our conditions, the electron transport and electron dynamics inside the target does not play a significant role in the ion acceleration process. This suggests that the dominant role in ion acceleration is played by processes at the target front, where laser energy is transferred to hot electrons in a similar way for all the investigated target thicknesses.

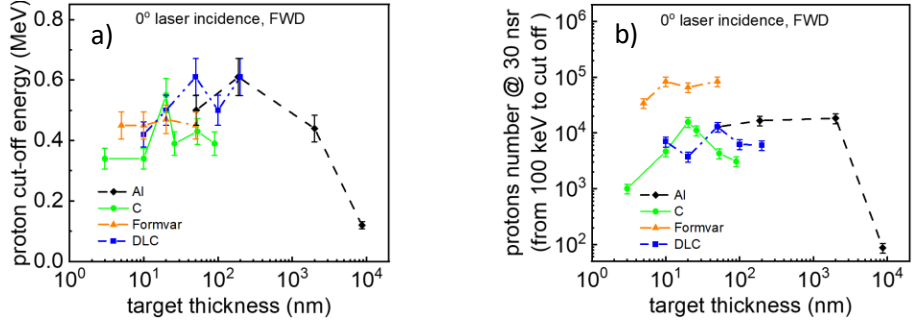

Fig. 4. a) The maximum energy of protons and b) the highest number of protons versus thickness emitted of the target in forward direction are shown at 0° laser incidence angle on the target for different target materials

Fig. 5 shows the emitted ion spectra in FWD for five successively increased thickness of $Al$ target at the same laser irradiance and at a 45° laser incidence angle on the target. An interesting phenomenon is that the carbon ions are mostly observed for the thinnest targets with dominance of $C^{4+}$. The ion charge-state density is gradually decreasing when target thickness is increased, while the maximum energy of ions stays virtually unchanged. For thicker targets, the number of carbon ions drops well below the detection limit (limited by the solid angle of reception of the Thomson spectrometer) and only protons are detected.

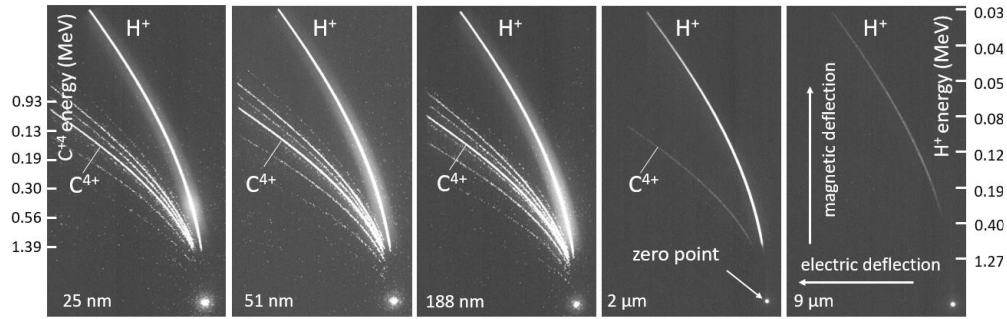

Fig. 5. The emitted ions spectral parabolic traces from $Al$ target measured in the FWD direction at four successively increasing thicknesses with the highest proton cut off energy.

In the spectra of BWD accelerated ions, such phenomenon was not observed. Carbon ions with dominance of $C^{4+}$ were present in all cases. At a 0° laser incidence angle it is difficult to make a clear statement since at this geometry, the particle number dropped by about an order of magnitude. Hence, the changes we may see could be related to the low number of detected particles.

IV. CONVERSION EFFICIENCY

From the measured number of particles in ion spectra (Fig. 3 b), d) and Fig 4b)) and the divergence of the proton beam about 5° [35], one can derive the spectral energy content in the proton beam. The assumption that the divergence of the proton beam is the same for all target thicknesses is rather conservative as the proton beam divergence may increase as target thickness decreases. As an example, Fig. 6 shows the spectral energy content in the proton beam for an $Al$ target, the one with the most systematic change of target thickness, both in FWD and BWD directions at 45° (Fig. 6 a) and b)), and in FWD direction at 0° angle of incidence (Fig. 6c)). It is noteworthy that at 45° laser incidence in FWD direction (Fig. 6a)) there is a plateau in the energy content in the beam with energies 0.2 MeV to 0.8 MeV for 51 nm, 188 nm and 2µm thick targets. For the thinnest (25 nm) and the thickest (9 µm) targets there is a clear maximum at around 0.2 MeV. The plateau was not observed in BWD at 45° (Fig. 6b)) nor FWD at 0° laser incidence (Fig. 6c)). Instead, the energy content in the beam had a maximum of about 0.2 MeV for all thicknesses besides the 9 µm target where the energy content was an order of magnitude lower (Fig. 6 b) and c)).

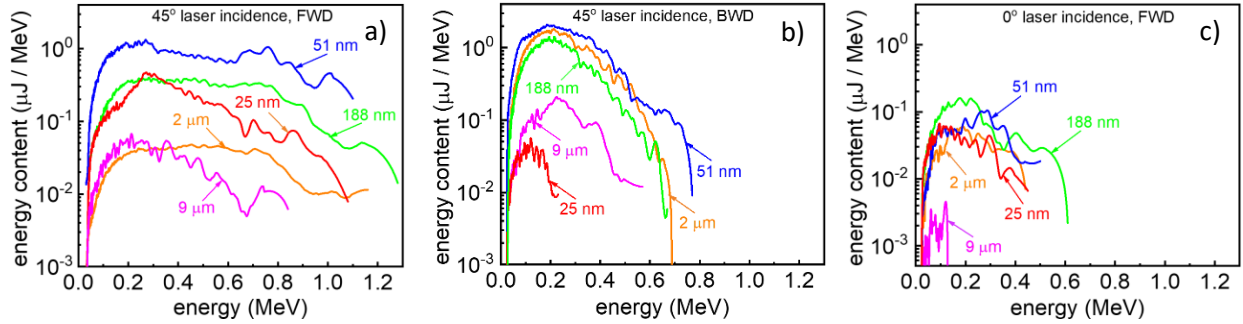

Fig. 6. The spectral energy content in the proton beam from *Al* targets at different thicknesses for FWB and BWD directions at 45° laser incidence and for FWD at 0° incidence.

On the whole, all this indicates a highly efficient transfer of laser energy to the target and acceleration of ions. On the basis of the energy content of the beam shown Fig. 6, a conservative estimate can be given for the conversion efficiency of laser energy to protons. Fig. 7 shows the conversion efficiency of laser energy into protons with energies above 100 keV to the cut-off for *Al* targets, depending on the target thicknesses. There is a clear maximum, reaching ~1.4% at a 51 nm thick target. However, the optimum target thickness can be given with some uncertainty due to the target thickness steps.

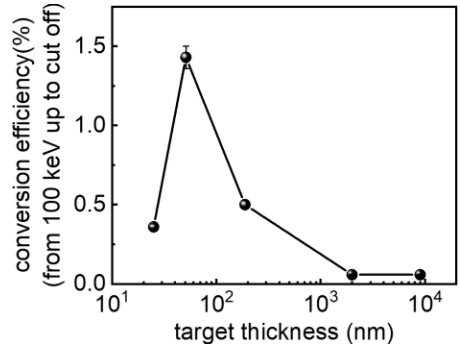

Fig. 7. Dependence of the conversion efficiency of laser energy into protons with energies above 100 keV on the thickness of *Al* targets in FWD direction at a 45° laser incidence angle (derived from Fig. 6).

It is known that target thickness should be properly matched to the laser intensity in order to obtain maximum ion energy. Earlier numerical modelling showed that protons can be accelerated to the highest energies at the targets of optimum thickness and density [9,10,11,13]. For target thickness at which relativistic self-induced transparency occurs near the laser peak [36], the protons acquire a high energy. In the experiments, the target thickness and laser pre-pulse found to be a key factors for high conversion efficiency ~4% at $10^{21}$ W/cm$^2$ [6], or, for optimization of the source properties [3]. Hence, the proper selection of foil target optimum thickness for certain laser parameters results in a very promising increase of the proton energy.

Ion acceleration from a hot plasma layer strongly depends on foil thickness $L$, and Debye radius of hot electrons $\lambda_{Dh}$. For $L \ll \lambda_{Dh}$, is the directed Coulomb explosion regime [11], where ions are accelerated from positively charged plasma layer. However, the ion energy and number are limited by the restricted amount of the charge involved in the interaction, considering the ultrathin target of this case. For $L > \lambda_{Dh}$, the ion acceleration takes place through the self-consistent electrostatic accelerating field generated by fast electrons escaping in a vacuum (so-called target normal sheath acceleration (TNSA) mechanism. Transition between the regimes happens at $L = L_M \leq \lambda_{Dh}$, where ion energy and number are maximal. We may estimate $L_M$ from the Debye length of hot electrons $\lambda_{Dh} = (\varepsilon_e^{(h)}/4\pi e^2 n_h)^{1/2}$, their density $n_h$, and average electron energy $\varepsilon_e^{(h)}$. We use the ponderomotive estimate for $\varepsilon_e^{(h)} = mc^2(\sqrt{1 + a^2/2} - 1)$ ~370 keV [37] and estimate $n_h$ from the approximate balance $\varepsilon_e^{(h)} n_h v_h \sim \rho I$, where $v_h$ is the characteristic electron velocity (close to speed of light): $v_h = c(\varepsilon^2 + 2\varepsilon)^{1/2}/(\varepsilon + 1)$, $\varepsilon = \varepsilon_e^{(h)}/mc^2$, or $\varepsilon(n_h/n_c)(v_h/c) \sim a^2/2$, $\rho$ is the laser energy absorption coefficient, $I$ is the laser intensity. The estimation for the characteristic hot electrons density reads as

$$\frac{n_h}{n_c} = \frac{a^2}{2} \frac{\varepsilon + 1}{\varepsilon\sqrt{2\varepsilon + \varepsilon^2}}$$

In our experiment $\rho \approx 20\% - 25\%$ [38], $a$ is close to 1.4, so $n_h \sim 1.8 n_c$, where the critical density $n_c = 1.7 \times 10^{21}$ cm$^{-3}$, so $L_M \leq \lambda_{Dh} \sim 115$ nm, in reasonable agreement with Fig.7.

Additionally, one can estimate the maximum proton energy from exploding plasma of density $\bar{n}_h$, with thickness $L \ll \lambda_{Dh}$, and transversal size $D$ is $\varepsilon_p \approx r_e L D \bar{n}_h m_e c^2$, where $r_e$ is the classical electron radius [11]. The averaged density over the large size ($\propto \lambda_{Dh}$) fulfilled by expelled electrons turns out to be $\lambda_{Dh}/l_{skin}$ times less than the density of locally heated electrons in the skin layer ($l_{skin}$). Then, the estimation of the maximum proton energy, $\varepsilon_p \approx r_e (L/\lambda_{Dh}) D l_{skin} n_h m_e c^2$, leads to the proton energy of the ~1 MeV level for the optimum thickness $L \simeq \lambda_{Dh}$.

V DISCUSSION

In laser energy transfer to the target and the resulting ion acceleration, the electron dynamics in the field of a relativistic laser pulse plays decisive role. This is determined by the spatiotemporal intensity distribution and polarisation of radiation. The parameters of the electron motion can be found by solving the Newton equation with the Lorentz force. However, to describe the motion of an electron in a high-frequency field is associated with significant difficulties due to the large number of field oscillations. Therefore, for many-cycle laser pulses the Lorentz force moving the electrons is averaged over high-frequency oscillations [37], called ponderomotive force. Radiation pressure in a plasma has an effects on the electrons via the ponderomotive force. Obviously, the concept of pondermotive force for a very short, few-cycle pulse needs to be reconsidered, and the motion of an electron must be analysed by using the exact Lorentz force. Therefore, the phenomena related to radiation pressure, like compression of the electron-density profile at the target front and related phenomena [39,40], cannot be applied here. Our efforts to prove the above with PIC simulations have revealed that a 3D model is needed with nm resolution. This is, however, hardly supported by available computing capacities. Hence, we present here a calculation that supports the major experimental findings on ion acceleration with few cycle relativistic laser pulses.

At an oblique angle of incidence of ultrashort- and ultrahigh-contrast laser pulses on a target, the laser energy is transferred to the target mainly due to the so-called vacuum heating via the Brunel's effect [41, 42] and $v \times B$ – heating [43]. The step-like plasma-vacuum interface prevents effective electron vacuum heating since a very high electron density of the solid target restricts the number of electrons involved in the interaction. However, one has to account for a plasma expansion from already $I \sim 10^{10}$ W/cm$^2$ intensity on the target at ~3.5 ps before the laser pulse intensity reaches to its maximum. With the expansion velocity $c_s \sim$ (1 - 2) $\times 10^4$ m/s [44], also calculated as $c_s = \sqrt{Z_i T_e / m_i}$, where $z_i = 3 - 4$, for hydro-carbon contaminant $m_i = 12 m_p$ and for $T_i = 20 - 30$ eV (somewhat higher than the minimal electron temperature equal to the quiver electron energy), during the time $\tau_{pp}$ ~3.5 ps, the plasma from the target front side extends at least to $L_{pp} \sim c_s \tau_{pp} \sim$ (30 – 60) nm. Thus, in the vicinity of the critical density, $n_c$, an exponential density profile with a characteristic gradient length $l \approx L_{pp} / \ln(n_{max}/n_c) \approx 5.5 - 11$ nm is formed. Such tiny preplasma itself does not directly affect the acceleration of ions by means of electron thermal pressure and cannot provide resonant absorption [41, 42]. It is due to the nonlinear effect of suppression of the plasma resonance field amplitude and "switching-off" of the resonant absorption with increasing laser intensity to the level $I \sim 10^{17}$ W/cm$^2$. This is a consequence of the relativistic nonlinearity of the electron plasma in the region of a critical density [45, 46]. However, the main laser field interacts directly with such electron preplasma which further expands during the laser pulse to sub-micron size through the self-consistent vacuum heating effect. Additionally, the laser energy coupling to the target depends on relativistic skin depth $l_{skin} = \gamma^{1/2} c / \omega_{ep}$.

Hence, we assume that the laser pulse produces high energy electrons through the Brunel's like heating from tiny preplasma and push them FWD to the rear of the target where build up charge separation field accelerates ions, (TNSA-like acceleration). It is important that classical Brunel's vacuum heating is not applicable because with Gaussian temporal shape of the laser pulse a preplasma extends in a vacuum to the considerable distance in comparison with a quarter laser wavelength. Therefore, it is likely that electrons experience stochastic heating in the complicated fields: incident and reflected p-polarized electromagnetic fields plus electrostatic plasma field [47].

Regarding the BWD accelerated ions at oblique laser incidence, one needs to take into account the strong charge separation in a tiny pre-plasma at the target front. The laser pulse pushes the electrons out of the focal spot domain forming a pancake-shaped charged cavity (similar to that discussed in Ref. [48]) in the pre-plasma with the averaged density, $<n_e>$. The cavity depth is equal to the skin depth $l_{skin}$, the transversal size $D$ is around the laser hot spot size, and $D \gg l_{skin}$. Being accelerated in the exploding Coulomb field of the charged cavity $E_c \approx 2\pi e l_{skin} <n_e>$ during its lifetime, $\tau_c$, which as a lowest estimate can be a duration of the laser pulse, $\tau_L$ ($\tau_c \cong \tau_L$), the protons gain energy $\varepsilon_p \approx (2/m_p)(\pi e^2 l_{skin} <n_e> \tau_c)^2 \approx 0.8$ MeV. The 0.8 MeV energy of the BWD accelerated protons, which is comparable to the energy of protons accelerated in FWD direction at thick targets, indicates an effective acceleration of ions by the charged cavity. More precisely, the duration of neutralization of a self-consistently expanding cavity (by ambient electrons) may last longer than the pulse duration $\tau_c \gtrsim \tau_L$ and, hence, increase the energy of the BWD accelerated protons.

It should be noted that as the target thickness increases, the resistive heating associated with the neutralisation return current - that is a result of a collective mechanism associated with the transfer of fast electrons [49,50,51] becomes important. Therefore, the electrons with an estimated typical energy $\varepsilon_e^{(h)} \approx 370$ keV have a stopping length which does not exceed 10 µm (due to depletion relevant to target heating by return current). As a result, the TNSA acceleration from the rear of the thick targets (above 10 µm) will be terminated. Nevertheless, at thick targets the measured cut-off energy of ~ 0.8 MeV both in FWD and BWD directions (see Fig. 3) indicates that particles are accelerated by the exploding Coulomb field of the charged cavity [47]. In this scenario, however, heavy ions (carbon) will also be symmetrically accelerated in both directions, but in the FWD direction they will be attenuated more than the protons in the target material and have less energy than BWD accelerated ions, i.e., the thick targets will be opaque to ions.

At normal laser incidence on the target the $v \times B$ heating dominates the energy transfer process over Brunel's absorption. For our moderate laser intensity, $a_0 \sim 2$, and very short pulse duration the generation of weakly relativistic hot electrons ($v < c$) at step-like plasma-vacuum interface prevents effective electron heating. Even Gaussian temporal shape of the laser pulse doesn't create even tiny electron preplasma while the electron oscillations are along the target surface. It leads to a very small field amplitude on the target surface in proportion to the smallness of the skin depth to the laser wavelength ratio $\sim \gamma^{1/2} \omega_0 / \omega_{ep}$, where $\omega_0$ and $\omega_{ep}$ are the laser field and electron plasma frequencies, respectively, $\gamma = \sqrt{1 + a_0^2/2}$ is the relativistic parameter, $a_0 = eE_0/m_e\omega_0 c$ is the normalized laser field amplitude, and $e$, $n_e$, $m_e$ are the electron charge, density, and mass, respectively. As a result the measured proton energies twice lower as compare with 45° laser incidence on the target. At oblique laser incidence the absorption is relevant to $p$-component of a laser field and its overlapping with short preplasma increases with the angle ($\theta$) of incidence as $L_{pp}/\cos\theta$, therefore one should expect an enhancement of proton acceleration with increase $\theta$ until almost grazing incidence. This will require further experimental verification for angles $\theta > 45°$.

V. SUMMARY

To summarize, ion acceleration experiments were carried out with few cycle relativistic laser pulses at intensities of about $10^{19}$ W/cm² and very high temporal contrast. The measurements of energies of ions accelerated from different target materials with different electron and hydrogen densities ($C$, DLC, formvar and $Al$ foil targets) and thicknesses ranging from 5 nm to 9 µm at laser incidence angles of 0∘ and 45∘ to the target normal have shown weak dependence on target thickness. From the measured number of particles, the spectral energy content of the proton beam was derived, showing a slowly decaying plateau over a wide energy range for forward accelerated protons at 45°angle of laser incidence. A conservative estimate of laser energy conversion efficiency to protons showed a clear maximum dependant on target thickness reaching ~1.4% at a 51 nm thick $Al$ target. Existence of an optimum thickness was explained with the transition between the major of ion acceleration schemes as from a directed Coulomb explosion to a self-consistent electrostatic field. In BWD direction the protons are accelerated in a Coulomb field of charged cavity with the lifetime mainly determined by the duration of the laser pulse. Despite the strong Coulomb field, the short acceleration time does not allow the protons to gain energy that exceeds the energy of protons accelerated FWD, as we saw in the experiments. However, the energy of protons accelerated BWD is comparable to that in FWD direction.

This is an interesting fact for our ultrahigh contrast laser–target interaction case compared to the usually observed considerable difference in the energies of FWD and BWD accelerated ions for similar temporal contrast [52]. The full process is more complex than our assumptions and it calls for 3D simulation with nm-resolution once the computing capacity is available.

As an outlook, the measured characteristics of FWD accelerated protons demonstrates that the use of CD instead of $CH$ would make it possible to accelerate deuterons to energies required for fusion neutron production from, e.g., the same CD-catcher. Once the targetry allows for the use of the kHz repetition rate SYLOS laser of ELI-ALPS, it would be a prototype for a 2.45 MeV neutron source. Additionally, implantation of protons with an energy of ~1 MeV generated by a high repetition rate laser can also be used in semiconductor technology as a new technique to obtain free-standing semiconductor films with thicknesses in the range of 10–20 µm [53].

ACKNOWLEDGEMENT

The authors are grateful for the fruitful discussions with Christos Kamperidis, Subhendu Kahaly, Martin Matys, Petr Valenta, Georg Korn, Sergei Bulanov, and Zsolt Fülöp. The authors are indebted to Arnold Farkas, Karoly Mogyorósi, Arpad Mohacsi, Rita Szabó, and Bence Kis for the technical help in the experiments. The project has been supported by the Hungarian Ministry of Technology and Innovation (contract # IFHÁT/1039-4/2019-ITM_SZERZ), the Hungarian National Research, Development, and Innovation Office through the National Laboratory program (contract # NKFIH-877-2/2020 and NKFIH-476-4/2021), and by the Ministry of Education, Youth and Sports of the Czech Republic (contract # CZ.02.1.01/0.0/0.0/16_019/0000789). The ELI-ALPS project is supported by the Hungarian Government and the European Regional Development Fund (contract # GINOP-2.3.6-15-2015-00001). V. Yu. B. acknowledges support from the Russian Foundation for Basic Research (Grant No. 20-21-00 023).

[1] F. Wagner, O. Deppert, C. Brabetz, P. Fiala, A. Kleinschmidt, P. Poth, V. A. Schanz, A. Tebartz, B. Zielbauer, M. Roth, T. Stöhlker, and V. Bagnoud, PRL 116, 205002 (2016)
[2] C.Danson, D. Hillier, N. Hopps, and D. Neely, High Power Laser Science and Engineering, 3, e3 (2015)
[3] M. Borghesi, A. Bigongiari, S. Kar, A Macchi, L. Romagnani, P. Audebert, J. Fuchs, T. Toncian, O. Willi, S V Bulanov, A J. Mackinnon and J. C. Gauthier Plasma Phys. Control. Fusion 50 124040 (2008).
[4] S. Steinke, A. Henig, M. Schnürer, T. Sokollik, P.V. Nickles, D. Jung, D. Kiefer, R. Hörlein, J. Schreiber, T. Tajima, X.Q. Yan, M. Hegelich, J. Meyer-ter-Vehn, W. Sandner and D. Habs Laser and Particle Beams 28, 215 (2010)
[5] A. Macchi, M. Borghesi, and M. Passoni, Rev. Mod. Phys. 85, 751 (2013)
[6] J. S. Green, A. P. L. Robinson, N. Booth, D. C. Carroll, R. J. Dance, R. J. Gray, D. A. MacLellan, P. McKenna, C. D. Murphy, D. Rusby, and L. Wilson Appl. Phys. Lett. 104, 214101 (2014)
[7] C. Scullion, D. Doria, L. Romagnani, A. Sgattoni, K. Naughton, D. R. Symes, P. McKenna, A. Macchi, M. Zepf, S. Kar, and M. Borghesi Phzs. Rev. Lett. 119, 054801 (2017)
[8] R. A. Snavely 1, M. H. Key, S. P. Hatchett, T. E. Cowan, M. Roth, T. W. Phillips, M. A. Stoyer, E. A. Henry, T. C. Sangster, M. S. Singh, S. C. Wilks, A. MacKinnon, A. Offenberger, D. M. Pennington, K. Yasuike, A. B. Langdon, B. F. Lasinski, J. Johnson, M. D. Perry and E. M. Campbell Phys. Rev. Lett. 85, 2945 (2000)
[9] E. d'Humieres, E. Lefebvre, L.Gremillet, and V.Malka Phys. Plasma, 12, 062704 (2005)
[10] T. Z. Esirkepov, M. Yamagiwa, and T. Tajima, Phys. Rev. Lett. 96, 105001 (2006)
[11] S.S.Bulanov, A.Brantov, V.Yu.Bychenkov, V.Chvykov, G.Kalinchenko, T.Matsuoka, P.Rousseau, S.Reed, V.Yanovsky, D.W.Litzenberg, K.Krushelnick, and A.Maksimchuk, Phys. Rev. E. 78, 026412 (2008).
[12] X.Q. Yan, T. Tajima, B.M. Hegelich, L. Yin, and D. Habs, Applied Phys. B 98, 711 (2010).
[13] E. d'Humieres, A. Brantov, V.Yu. Bychenkov, and V. T. Tikhonchuk, Phys. Plasmas 20, 023103 (2013)

[14] M. Borghesi, D. H. Campbell, A. Schiavi, M. G. Haines, O. Willi, A. J. MacKinnon, P. Patel, L. A. Gizzi, M. Galimberti, R. J. Clarke, F. Pegoraro, H. Ruhl, and S. Bulanov Phys. Plasmas 9, 2214 (2002).

[15] S. Kar, A. Green, H. Ahmed, A. Alejo, A. P. L. Robinson, M. Cerchez, R. Clarke, D. Doria, S. Dorkings, J. Fernandez, S. R. Mirfayzi, P. McKenna, K. Naughton, D. Neely, P. Norreys, C. Peth, H. Powell, J. A. Ruiz, J. Swain, O. Willi, and M. Borghesi New J. Phys. 18, 053002 (2016).

[16] A. Pelka, G. Gregori, D. O. Gericke, J. Vorberger, S. H. Glenzer, M. M. Gunther, K. Harres, R. Heathcote, A. L. Kritcher, N. L. Kugland, B. Li, M. Makita, J. Mithen, D. Neely, C. Niemann, A. Otten, D. Riley, G. Schaumann, M. Schollmeier, A. Tauschwitz, and M. Roth Phys. Rev. Lett. 105, 265701 (2010).

[17] S. D. Kraft, C. Richter, K. Zeil, M. Baumann, E. Beyreuther, S. Bock, M. Bussmann, T.E. Cowan, Y. Dammene, W.Enghardt, U. Helbig, L. Karsch, T. Kluge, L. Laschinsky, E. Lessmann, J. Metzkes, D. Naumburger, R. Sauerbrey, M. Schürer, M. Sobiella, J. Woithe, U. Schramm and J. Pawelke, New J. Phys. 12, 085003 (2010).

[18] S. Raschke, S. Spickermann, T. Toncian, M. Swantusch, J. Boeker, U. Giesen, G. Iliakis, O. Willi, and F. Boege Sci. Rep. 6, 32441 (2016).

[19] T. Ziegler, D. Albach, C. Bernert, S. Bock, F.-E. Brack, T. E. Cowan, N. P. Dover, M. Garten, L. Gaus, R. Gebhardt, I. Goethel, U. Helbig, A. Irman, H. Kiriyama, T. Kluge, A. Kon, S. Kraft, F. Kroll, M. Loeser, J. Metzkes-Ng, M. Nishiuchi, L. Obst-Huebl, T. Püschel, M. Rehwald, H.-P. Schlenvoigt, U. Schramm and K. Zeil Sci.Rep. 11, 7338 (2021)

[20] A. Permogorov, G. Cantono, D. Guenot, A. Persson, and Claes-Göran Wahlström, Sci. Rep. 12, 3031 (2022)

[21] Jianhui Bin, Lieselotte Obst-Huebl, Jian-Hua Mao, Kei Nakamura, Laura D. Geulig, Hang Chang, Qing Ji, Li He, Jared De Chant, Zachary Kober, Anthony J. Gonsalves, Stepan Bulanov, Susan E. Celniker, Carl B. Schroeder, Cameron G. R. Geddes, Eric Esarey, Blake A. Simmons, Thomas Schenkel, Eleanor A. Blakely, Sven Steinke and Antoine M. Snijders Sci. Rep. 12, 1484 (2022)

[22] Cirrone GAP Cuttone G, Raffaele L, Salamone V, Avitabile T, Privitera G, Spatola C, Amico AG, Larosa G, Leanza R, Margarone D, Milluzzo G, Patti V, Petringa G, Romano F, Russo A, Russo A, Sabini MG, Scuderi V, Schillaci F, Valastro and LM, Front. Phys. 8, 56490 (2020);

[23] Florian Kroll, Florian-Emanuel Brack, Constantin Bernert, Stefan Bock, Elisabeth Bodenstein, Kerstin Brüchner, Thomas E. Cowan, Lennart Gaus, René Gebhardt, Uwe Helbig, Leonhard Karsch, Thomas Kluge, Stephan Kraft, Mechthild Krause, Elisabeth Lessmann, Umar Masood, Sebastian Meister, Josefine Metzkes-Ng, Alexej Nossula, Jörg Pawelke, Jens Pietzsch, Thomas Püschel, Marvin Reimold, Martin Rehwald, Christian Richter, Hans-Peter Schlenvoigt, Ulrich Schramm, Marvin E. P. Umlandt, Tim Ziegler, Karl Zeil and Elke Beyreuther, Nat. Phys. 18, 316 (2022).

[24] Sz. Toth, T. Stanislauskas, I. Balciunas, R. Budriunas, J. Adamonis, R. Danilevicius, K. Viskontas, D. Lengvinas, G. Veitas, D. Gadonas, A. Varanavičius, J. Csontos, T. Somoskoi, L. Toth, A. Borzsonyi and K. Osvay J. Phys. Photonics, 2, 045003 (2020).

[25] J.T. Morrison, S. Feister, K. D. Frische, D. R. Austin, G. K. Ngirmang, N. R. Murphy, C. Orban, E. A. Chowdhury, and W M Roquemore, New J. Phys. 20, 022001 (2018)

[26] MultiScan 3D project: https://cordis.europa.eu/project/id/101020100

[27] Gesellschaft für Schwerionenforschung (GSI). Laser Ion Generation, Handling and Transport (LIGHT) at GSI. URL: https://www.gsi.de/work/forschung /appamml/plasmaphysikphelix/experimente/light.htm.

[28] S. Fritzler, V. Malka, G. Grillon, J. P. Rousseau, F. Burgy, E. Lefebvre, E. d'Humiè`res, P. McKenna and K. W. D. Ledingham Appl. Phys. Lett. 83, 3039 (2003);

[29] E. Lefebvre, J. Appl. Phys. 100, 113308 (2006).

[30] M. Veltcheva, A. Borot, C. Thaury, A. Malvache, E. Lefebvre, A. Flacco, R. Lopez-Martens, and V. Malka, Phys. Rev. Lett. 108, 075004 (2012)

[31] D. Levy, I. A. Andriyash, S. Haessler, J. Kaur, M. Ouill´e, A. Flacco, E. Kroupp, V. Malka, and R. Lopez-Martens, Phys. Rev. Acc. Beams 25, 093402 (2022)

[32] P. K. Singh, K. F. Kakolee, T. W. Jeong, and S. Ter-Avetisyan Nucl. Instrum. Meth. A 829, 363 (2016)

[33] P. Varmazyar, P.K. Singh, Z. Elekes, Z. Halasz, B. Nagy, J.-G. Son, at al., Rev. Sci. Instrum. 93, 073301 (2022).

[34] T. Ceccotti, A. Le´vy, H. Popescu, F. Re´au, P. D'Oliveira, P. Monot, J. P. Geindre, E. Lefebvre, and Ph. Martin Phys. Rev. Lett. 39, 284 (1977).

[35] P.K. Singh, P.Varmazyar, B.Nagy, J.-G. Son, S. Ter-Avetisyan, and K. Osvay, Sci. Rep. 12, 8100 (2022).

[36] A. V. Brantov, E. A. Govras, and V. Yu. Bychenkov Phys, Rev, ST, 18, 021301 (2015).

[37] B. Quesnel, and P. Mora Phys. Rev. E, 58, 3719 (1998)

[38] Luca Fedeli, Arianna Formenti, Lorenzo Cialf, Andrea Pazzaglia and Matteo Passoni Sci. Rep. 8, 3834 (2018).

[39] V. A. Vshivkov Phys. Plasmas 5, 2727 (1998)

[40] F. Cattani, A. Kim, D. Anderson, and M. Lisak Phys. Rev. E 62, 1234 (2000).

[41] D. W. Forslund, J. M. Kindel, and K. Lee Phys. Rev. Lett. 39, 284 (1977).
[42] F. Brunel Phys. Rev. Lett. 59, 52 (1987).
[43] W. L. Kruer and K. Estabrook Phys. Fluids 28, 430 (1985).
[44] D. Batani, R. Jafer, M. Veltcheva, R. Dezulian, O. Lundh, F. Lindau, A. Persson, K. Osvay, C.-G. Wahlström, D. C. Carroll, P. McKenna, A. Flacco, and V. Malka, New J. Phys. 12, 045018 (2010).
[45] I. I. Metelskii, V. F, Kovalev and V. Yu. Bychenkov J. Exp. Theor. Phys. 133 236 (2021).
[46] S. K. Rajouria, K. K. Magesh Kumar, and V. K. Tripathi, Phys. Plasmas 20, 083112 (2013).
[47] Y. Sentoku y. V.Y. Bychenkov, K. Flippo, A. Maksimchuk, K. Mima, G. Mourou, Z.M. Sheng and D. Umstadte Appl. Phys. B 74, 207 (2002).
[48] V.Yu. Bychenkov, P.K. Singh, H. Ahmed, K.F. Kakolee, C. Scullion, T.W. Jeong, P. Hadjisolomou, A. Alejo, S. Kar, M. Borghesi, and S. Ter-Avetisyan Phys. Plasmas 24, 010704 (2017)
[49] M. Passoni, V. T. Tikhonchuk, M. Lontano, and V. Yu. Bychenkov Phys. Rev. E 69, 026411 (2004).
[50] J J Santos, A Debayle, Ph Nicolaï, V Tikhonchuk, M Manclossi, D Batani, A GuemnieTafo, J Faure, V Malka, and J J Honrubia J. Phys.: Conf. Ser. 112 022088 (2008)
[51] M. Sherlock, E. G. Hill, R. G. Evans, S. J. Rose, and W. Rozmus Phys. Rev. Lett. 113, 255001 (2014)
[52] S. Ter-Avetisyan, P.K. Singh, K.F. Kakolee, H. Ahmed, T.W. Jeong, C. Scullion, P. Hadjisolomou, M. Borghesi, and V. Yu. Bychenkov Nucl. Instrum. Methods A 909, 156 (2018).
[53] Kai Huang, Tiangui You, Qi Jia, Ailun Yi, Shibin Zhang, Runchun Zhang, Jiajie Lin, Min Zhou, Wenjie Yu, Bo Zhang, Xin Ou and Xi Wang Appl. Phys. A 124, 118 (2018).